\documentclass[prd,nofootinbib,amsfonts,notitlepage]{revtex4-1}
\usepackage[utf8]{inputenc}
\usepackage[T1]{fontenc}
\usepackage{hyperref}
\usepackage{graphicx,bm,amsmath,color}
\usepackage{bbold}
\usepackage{wasysym}
\usepackage{verbatim}
\usepackage{amssymb}

\usepackage{epstopdf}
\usepackage{animate}
\usepackage{xmpmulti}
\usepackage{centernot}
\usepackage{multirow}
\usepackage{tikz}
\usepackage{graphicx}
\usepackage{float}
\usepackage{mathtools}
\usepackage{comment}

\DeclareMathOperator{\arcsec}{arcsec}
\makeatletter

\newcommand{\be}{\begin{equation}}
\newcommand{\ee}{\end{equation}}
\newcommand{\beq}{\begin{eqnarray}}
\newcommand{\eeq}{\end{eqnarray}}
\newcommand{\ba}{\begin{align}}
\newcommand{\ea}{\end{align}}

\begin{document}

\title{Geometry of multi-particle systems with a relativistic deformed kinematics and the relative locality principle}
\author{J.J. Relancio}
\affiliation{Dipartimento di Fisica ``Ettore Pancini'', Università di Napoli Federico II, Napoli, Italy;\\
INFN, Sezione di Napoli, Italy;\\
Centro de Astropartículas y Física de Altas Energías (CAPA), Universidad de Zaragoza, Zaragoza 50009, Spain}
\email{relancio@unizar.es}

\begin{abstract}
There is a vast literature showing the connection between a deformed relativistic kinematics and a curved momentum space, and, in particular, how the former can be obtained from the geometrical properties of the latter. However, there is not any mention about the geometry of a multi-particle system making manifest a possible modification in the metric of one particle due to the presence of others. In this work we explore how a curved momentum metric depending on the particles involved in an interaction arises when considering a process. We also show that the principle of relative locality obtained in doubly special relativity from an action is achieved in this geometrical framework in a direct fashion. Moreover, this formalism allows us to generalize this principle when a curvature of spacetime is present in a natural way. Furthermore this geometrical setup allows us  to define a new momentum dependent space-time coordinates for a multi-particle system  in which locality of interactions is recovered.
\end{abstract}

\maketitle

\section{Introduction}
One of the challenges of theoretical physics nowadays is the formulation of a quantum gravity theory (QGT). The main difficulty in merging quantum field theory (QFT) and general relativity (GR) is the different role that spacetime plays in these theories: it is a static frame in QFT and a dynamical entity in GR. One would expect that the classical notion of spacetime should change when considering small scales. A completely new structure would arise, leading to a ``quantum'' spacetime, with novel properties we are not able to imagine nowadays.

This is a matter of study in the last years. For example, in loop quantum gravity~\cite{Sahlmann:2010zf,Dupuis:2012yw}, such structure takes the form of a spin foam~\cite{Wheeler:1955zz,Rovelli:2002vp,Ng:2011rn,Perez:2012wv}, that can be interpreted as a ``quantum'' spacetime, and in causal set theory~\cite{Wallden:2013kka,Wallden:2010sh,Henson:2006kf} and string theory~\cite{Mukhi:2011zz,Aharony:1999ks,Dienes:1996du}, non-locality effects appear~\cite{Belenchia:2014fda,Belenchia:2015ake}. This differs completely from the notion of Einstein spacetime~\cite{Einstein1905}, which is constructed via  the exchange of light signals. Moreover, if there is a  ``quantum'' spacetime, the propagation of massless particles with different energies could vary, for example, through an energy dependent velocity, and Einstein's construction looses sense. Furthermore, the description proposed by Einstein is useless when non-locality effects arise. 

However, the aforementioned theories are not fully satisfactory in the sense that they do not have well defined testable predictions, which might serve us as a guidance in building a theory of quantum gravity. So, instead of starting from a possible fundamental QGT, one could consider a different approach starting by its low energy limit. 

Since the structure of the spacetime should change for high energies, also its usual symmetries of it would be different. Regarding Lorentz invariance, there are two main scenarios: one can consider that for high energies a Lorentz invariance violation (LIV)~\cite{Colladay:1998fq,Kostelecky:2008ts} can arise, or that this symmetry is deformed, leading to the theories known as deformed special relativity (DSR)~\cite{AmelinoCamelia:2008qg}. These theories are not formulated as a QGT but as its possible a low energy, allowing us to explore possible phenomenological implications which might serve us as a guidance in building a  fundamental theory of quantum gravity.

LIV scenarios modify the kinematics of special relativity (SR) with the introduction of a deformed dispersion relation. New terms proportional to the inverse of a high-energy scale (normally considered to be the Planck scale) are added to the usual quadratic expression of SR. However, in DSR theories, besides the possibility of a deformed dispersion relation,\footnote{This ingredient is not an essential element in the kinematics of DSR. For example, in the so-called ``classical basis'' of $\kappa$-Poincaré~\cite{Borowiec2010}, the dispersion relation is the usual one of SR.} there are some deformed conservation laws for the momenta implying that the total momentum of a system of two (or more) particles is not derivable as the trivial sum of the initial momenta as in SR, but involves instead additional terms depending on both momenta and on the high-energy scale. Furthermore, in order to have a relativity principle present in these theories, it is mandatory to have deformed Lorentz transformations for the multi-particle system, making the deformed dispersion relation compatible with the conservation laws.

As commented above, a ``quantum'' spacetime could induce some unexpected features in particle kinematics; for example, the velocity of massless particles could depend on their energy, or  the interactions become  non-local. The first possible effect can be parametrized by a deformed dispersion relation, an ingredient that appears in both LIV and DSR scenarios. However, the lack of locality of interactions is only present in DSR theories due to the deformation of the conservation law of momenta, an effect known as relative locality~\cite{AmelinoCamelia:2011bm}. This can be easily understood just regarding the momentum as the generators of translations in spacetime: since the total momentum is a nonlinear function of the momenta of the particles, translations are different for each particle involved in the process.

In~\cite{Carmona:2019fwf} it was rigorously shown that all the ingredients of a relativistic deformed kinematics can be obtained from a maximally symmetric momentum space. In particular,   $\kappa$-Poincaré kinematics can be obtained identifying the isometries (translations and Lorentz isometries) and the squared distance of the metric with the deformed composition law, deformed Lorentz transformations and deformed dispersion relation respectively (the last two facts were previously contemplated in Refs.~\cite{AmelinoCamelia:2011bm,Lobo:2016blj}). In~\cite{Relancio:2020zok} the proposal of~\cite{Carmona:2019fwf} was generalized so allowing the metric to describe a curved spacetime, i.e. a generalization of GR including a deformed relativistic kinematics, which leads to a metric in the cotangent bundle depending on all the phase-space variables. This is a generalization of previous works in the literature, in which a metric that depends on the velocities (Finsler geometries)~\cite{Girelli:2006fw,Amelino-Camelia:2014rga,Lobo:2016xzq} and momenta (Hamilton geometries)~\cite{Barcaroli:2015xda,Barcaroli:2016yrl,Barcaroli:2017gvg} were regarded.

Following the relative locality idea, the study of the propagation and interaction of particles considering a curvature in both momentum and space-time spaces was carried out in Ref.~\cite{Cianfrani:2014fia}. In that paper an action with some nonlocal variables (defined by the space-time tetrad) is considered (differing from the approach of the aforementioned works in which the space-time coordinates are the canonical conjugated variables of the momentum), allowing them to generalize the relative locality action~\cite{AmelinoCamelia:2011bm} when a curvature in spacetime is present. 

In another vein, in some recent works~\cite{Carmona:2017cry,Carmona:2019vsh} it has been shown that one can construct some noncommutative coordinates in a two-particle system in such a way that in these coordinates one recovers locality of interactions. In~\cite{Carmona:2021gbg} the relationship between a curved momentum space and these generalized coordinates (in the one particle system) was explored. In particular, it was shown that the functions characterizing the noncommutative coordinates can be identified with the (inverse of the) tetrad in a momentum space metric. 

In this work, we will see how one can describe an interaction in the geometrical approach of~\cite{Relancio:2020zok,Relancio:2020rys}, from which we can deduce the relative locality principle. This  forces us to consider a metric for the phase space of two particles, depending in general on all the momenta involved in the process. Moreover, we are able to establish a relationship between this momentum metric and the noncommutative coordinates of a two-particle system in which the interactions are local.

The structure of the paper is as follows. We start by explaining  how the principle of relative locality arises from geometrical considerations for a flat spacetime in Sec.~\ref{sec:rl_flat}. In this section, we see that this defines a momentum metric depending on all the momenta involved in the interaction. In Sec.~\ref{sec:st}, we observe that the construction of a metric in the phase space of a two-particle system leads to a definition of spacetime in which interactions are local. We will apply this model to $\kappa$-Poincaré kinematics in Sec.~\ref{sec:kappa}, showing how it is possible to extend this work to a system of more than two particles, and to the most general kinematics at first order in a Taylor expansion on the high-energy scale deforming the kinematics in Sec.~\ref{sec:general}. In Sec.~\ref{sec:rl_curved} we show how to generalize the relative locality principle for a curved spacetime.  Finally, we see the conclusions in Sec.~\ref{sec:conclusions}.

\section{Relative locality in flat spacetime}
\label{sec:rl_flat}

In this section we start by resuming the original proposal of relative locality from the following action~\cite{AmelinoCamelia:2011bm} 
\begin{align}
S^{(2)} \,=&\, \int_{-\infty}^0 d\tau \sum_{i=1,2} \left[x_{-(i)}^\mu(\tau) \dot{p}_\mu^{-(i)}(\tau) + N_{-(i)}(\tau) \left[C(p^{-(i)}(\tau)) - m_{-(i)}^2\right]\right] \nonumber \\   
& + \int^{\infty}_0 d\tau \sum_{j=1,2} \left[x_{+(j)}^\mu(\tau) \dot{p}_\mu^{+(j)}(\tau) + N_{+(j)}(\tau) \left[C(p^{+(j)}(\tau)) - m_{+(j)}^2\right]\right] \nonumber \\
& + \xi^\mu (0)\left[{\cal P}^+_\mu(0) - {\cal P}^-_\mu(0)\right]\,,
\label{S2}
\end{align}
where $\dot{a}\doteq (da/d\tau)$ is a derivative of the variable $a$ with respect to the parameter $\tau$ along the trajectory of the particle, $x_{-(i)}$ ($x_{+(j)}$) are the space-time coordinates of the in-state (out-state) particles, $p^{-(i)}$ ($p^{+(j)}$) their four-momenta, $m_{-(i)}$ ($m_{+(j)}$) their masses, ${\cal P}^-$ (${\cal P}^+$) the total four-momentum of the in-state (out-state) defining the deformed composition law, $C(k)$ the function of a four-momentum $k$ defining the deformed dispersion relation, $\xi^\mu(0)$ are Lagrange multipliers that implement the energy-momentum conservation in the interaction, and $N_{-(i)}$ ($N_{+(j)}$) are Lagrange multipliers implementing the dispersion relation of in-state (out-state) particles.
 
Applying the variational principle to the action (\ref{S2}), one obtains the end (starting) space-time coordinates of the trajectories of the in-state (out-state) particles
\be
x_{-(i)}^\mu(0) \,=\, \xi^\nu(0) \frac{\partial {\cal P}^-_\nu}{\partial p^{-(i)}_\mu}(0)\,, \quad\quad\quad
x_{+(j)}^\mu(0) \,=\, \xi^\nu(0) \frac{\partial {\cal P}^+_\nu}{\partial p^{+(j)}_\mu}(0)\,.
\ee

We see from the previous equation that only an observer placed at the interaction point ($\xi^\mu(0)=0$) will see the interaction as local (all $x^\mu_J(0)$ coincide, being zero). One can choose the Lagrange multiplier $\xi^\mu(0)$ so the interaction will be local only for one observer, but any other one will see the interaction as non-local. This shows the loss of absolute locality, effect baptized as relative locality.

\subsection{Relative locality from geometry: the right particle}
We are going to obtain the relative locality principle from the geometrical approach of a metric in the cotangent bundle used in~\cite{Relancio:2020zok,Relancio:2020rys}. In~\cite{2012arXiv1203.4101M} a line element in the cotangent bundle is defined as  
\begin{equation}
\mathcal{G}\,=\, g_{\mu\nu}(x,k) dx^\mu dx^\nu+g^{\mu\nu}(x,k) \delta k_\mu \delta k_\nu\,,
\label{eq:line_element_ps} 
\end{equation}
where 
\begin{equation}
\delta k_\mu \,=\, d k_\mu - N_{\nu\mu}(x,k)\,dx^\nu\,. 
\end{equation}
Here, $N_{ \nu \mu}$ are the so-called coefficients of the nonlinear connection. In GR, the coefficients of the nonlinear connection are given by
\begin{equation}
N_{\mu\nu}(x,k)\, = \, k_\rho \Gamma^\rho_{\mu\nu}(x)\,,
\label{eq:nonlinear_connection}
\end{equation} 
where $ \Gamma^\rho_{\mu\nu}(x)$ is the affine connection. Then, when the metric is such that it does not depend on the space-time coordinates, these coefficients vanish, making that~\eqref{eq:line_element_ps} becomes
\begin{equation}
\mathcal{G}\,=\, g_{\mu\nu}(k) dx^\mu dx^\nu+g^{\mu\nu}(k) d k_\mu d k_\nu\,.
\label{eq:line_element_ps_flat} 
\end{equation}
It is important to note that $x$ and $k$ are canonically conjugated variables, having then the usual structure of Poisson brackets 
\begin{equation}
\lbrace{k_\nu, x^\mu \rbrace}\,=\,\delta^\mu_\nu\,.
\end{equation}

As we commented in the introduction, the composition law is defined as the isometries of the momentum metric, i.e~\cite{Carmona:2019fwf} 
\begin{equation}
g^{\mu\nu}(q) d q_\mu d q_\nu\,=\, g^{\mu\nu}\left(p\oplus q\right)d \left(p\oplus q\right)_\mu d \left(p\oplus q\right)_\nu\,,
\label{eq:line_element_isometry} 
\end{equation}
which leads to
\begin{equation}
g_{\mu\nu}\left(p\oplus q\right) \,=\,\frac{\partial \left(p\oplus q\right)_\mu}{\partial q_\rho} g_{\rho\sigma}(q)\frac{\partial \left(p\oplus q\right)_\nu}{\partial q_\sigma}\,.
\label{eq:composition_isometry}
\end{equation} 

Then, we can apply such transformation to the line element of Eq.~\eqref{eq:line_element_ps_flat} for the flat space-time case for the sake of simplicity (the curved space-time case will be considered in Sec.~\ref{sec:rl_curved}). Since it is an isometry, we would have in principle
\begin{equation}
\mathcal{G}\,=\, g_{\mu\nu}(k) dx^\mu dx^\nu+g^{\mu\nu}(k) d k_\mu d k_\nu\,=\,g_{\mu\nu}(\epsilon \oplus k) dx^\mu dx^\nu+g^{\mu\nu}(\epsilon\oplus k) d (\epsilon \oplus k)_\mu d (\epsilon \oplus k)_\nu\,,
\label{eq:line_element_ps_1} 
\end{equation}
being $\epsilon$ the parameter of the translation. However, it is easy to see that the previous equation cannot hold due to Eq.~\eqref{eq:composition_isometry}. This means that the composition law is only an isometry for the vertical line element but not for the whole line element, and in particular, not for the space-time line element. Then, we need to consider that, in order to have an isometry of the complete phase-space line element, the space-time coordinates changes when applying a momentum translation 
\begin{equation}
\mathcal{G}\,=\, g_{\mu\nu}(k) dx^\mu dx^\nu+g^{\mu\nu}(k) d k_\mu d k_\nu\,=\,g_{\mu\nu}(\epsilon \oplus k) d\xi^\mu d\xi^\nu+g^{\mu\nu}(\epsilon\oplus k) d (\epsilon \oplus k)_\mu d (\epsilon \oplus k)_\nu\,,
\label{eq:line_element_ps_2} 
\end{equation}

Due to the relationship of Eq.~\eqref{eq:composition_isometry} involving the composition law and the metric, one can find the following differential equation
\begin{equation}
\frac{\partial x^\mu }{\partial \xi^\rho}\,=\,\frac{\partial \left(\epsilon \oplus k\right)_\rho}{\partial k_\mu} \,.
\end{equation}
Then, one can solve it finding that 
\begin{equation}
x^\mu\,=\,\frac{\partial \left(\epsilon \oplus k\right)_\rho}{\partial k_\mu}\xi^\rho+\text{const}\,.
\end{equation}
Without loss of generality one can set the constant to be zero and then, one can particularize the previous equation for a given phase-space point $(x(0),k(0))$, obtaining
\begin{equation}
x^\mu(0)\,=\,\frac{\partial \left(\epsilon \oplus k(0)\right)_\rho}{\partial k_\mu(0)}\xi^\rho(0)\,.
\label{eq_rel_loc_proof}
\end{equation}
This shows that under an isometry in momentum space the space-time coordinates must also change in order to have an isometry for the whole phase space. 

Now, we are going to study what happens with space-time coordinates when an interaction takes place. We start by considering a $2-2$ scattering process (for simplicity) with two incoming particles with phase-space coordinates $\left(y,p\right)$ and $\left(z,q\right)$ and two outgoing particles with phase-space coordinates $\left(u,k\right)$ and $\left(w,l\right)$. We consider that the total momentum of the system of these two particles before and after the interaction is given by
\begin{equation}
\left(p(0)\oplus q(0)\right)_\mu\,=\,\left(k(0)\oplus l(0)\right)_\mu\,,
\label{eq:conservation_composition}
\end{equation}
where $0$ is representing the value of the momentum when the interaction takes place. This total momentum is conserved through a nonlinear composition law.

We firstly consider that the particle with momentum $q$ in the initial state corresponds to the one with momentum $l$ after the interaction. This is a very particular case used in order to illustrate the feature of relative locality. As we will see in Sec.~\ref{sec:rl_left_particle}, this simple implementation leads to some problems for the left particle, forcing us to consider a more general scenario in Sec.~\ref{sec:line_element}. 

The crucial assumption we are going to consider is that, in an interaction, the initial and final points for each particle are depicted by isometries in phase space of the kind of Eq.~\eqref{eq:line_element_ps_2}. Then, the momentum of each particle changes through the momentum composition law (defined as isometry in momentum space), and the space-time coordinates changes according to it, as we saw in Eq.~\eqref{eq_rel_loc_proof}. Then, the phase-space line element for the right particle (with momentum $q$ before the interaction and $l$ after it) is 
\begin{equation}
g_{\mu\nu}(q) dz^\mu dz^\nu+g^{\mu\nu}(q) d q_\mu d q_\nu\,=\,g_{\mu\nu}(l) dw^{\mu} dw^{\nu}+g^{\mu\nu}(l) d l_\mu d l_\nu\,.
\end{equation}

In order to make things easier, we can define an intermediate state with phase-space coordinates $(\xi, \left(p\oplus q\right))$ between the previous line elements, making that  
\begin{equation}
\begin{split}
&g_{\mu\nu}(q) dz^\mu dz^\nu+g^{\mu\nu}(q) d q_\mu d q_\nu\,=\,
g_{\mu\nu}\left(p\oplus q\right) d\xi^\mu d\xi^\nu+g^{\mu\nu}\left(p\oplus q\right) d \left(p\oplus q\right)_\mu d\left(p\oplus q\right)_\nu\,=\,\\ 
&g_{\mu\nu}\left(k\oplus l\right) d\xi^\mu d\xi^\nu+g^{\mu\nu}\left(k\oplus l\right) d \left(k\oplus l\right)_\mu d\left(k\oplus l\right) _\nu\,=\,
 g_{\mu\nu}(l) dw^{\mu} dw^{\nu}+g^{\mu\nu}(l) d l_\mu d l_\nu\,,
\end{split}
\label{eq:line_vertex}
\end{equation}
where $\xi$ and  $(p\oplus q)$ are canonical conjugated variables and also $\xi$ and  $(k\oplus l)$, since Eq.~\eqref{eq:conservation_composition} holds. 

Applying the same procedure of Eq.~\eqref{eq_rel_loc_proof} we can relate the coordinates $z^\mu$ with $\xi^\mu$: if  $\xi^\rho(0)$ is the vertex of the interaction, i.e. a given particular coordinate defining where the interaction takes place, then the  $z^\mu(0)$ coordinates of the right-ingoing particle corresponding to such vertex is 
\begin{equation}
z^\mu(0)\,=\,\frac{\partial \left(p(0)\oplus q(0)\right)_\rho}{\partial q_\mu(0)}\xi^\rho(0)\,,
\label{eq_rel_loc_r}
\end{equation}
and in a similar way for the right-outgoing particle coordinates after the interaction
\begin{equation}
w^{\mu}(0)\,=\,\frac{\partial \left(k(0)\oplus l(0)\right)_\rho}{\partial l_\mu(0)}\xi^\rho(0)\,.
\end{equation}
This is the same result obtained in~\cite{AmelinoCamelia:2011bm} for one of the particles. We are going to see that for the other particle involved in the interaction this result cannot be obtained in a direct way. 

\subsection{Relative locality from geometry: the left particle}
\label{sec:rl_left_particle}
One could naively look for a similar derivation of the relative locality principle for the left particle, imposing 
\begin{equation}
\begin{split}
&g_{\mu\nu}(p) dy^\mu dy^\nu+g^{\mu\nu}(p) d p_\mu d p_\nu\,=\,g_{\mu\nu}\left(p\oplus q\right) d\xi^\mu d\xi^\nu+g^{\mu\nu}\left(p\oplus q\right) d \left(p\oplus q\right)_\mu d\left(p\oplus q\right)_\nu\,=\,\\ 
&g_{\mu\nu}\left(k\oplus l\right) d\xi^\mu d\xi^\nu+g^{\mu\nu}\left(k\oplus l\right) d \left(k\oplus l\right)_\mu d\left(k\oplus l\right) _\nu\,=\, g_{\mu\nu}(k) du^{\mu} du^{\nu}+g^{\mu\nu}(k) d k_\mu d k_\nu\,.
\end{split}
\end{equation}
However, in order to obtain the relation between the $z^\mu$ and $\xi^\mu$ we have used Eq.~\eqref{eq:line_element_isometry}, making use explicitly of the isometry condition of the composition law~\eqref{eq:composition_isometry}. A similar equation for the momentum of the left particle cannot be addressed since  
\begin{equation}
g_{\mu\nu}\left(p\oplus q\right) \,\neq\,\frac{\partial \left(p\oplus q\right)_\mu}{\partial p_\rho} g_{\rho\sigma}(p)\frac{\partial \left(p\oplus q\right)_\nu}{\partial p_\sigma}\,,
\label{eq:composition_isometry_false}
\end{equation} 
which does not hold due to the non-symmetricity of the composition law. This impedes us to obtain the expected result
\begin{equation}
y^\mu(0)\,=\,\frac{\partial \left(p(0)\oplus q(0)\right)_\rho}{\partial p_\mu(0)}\xi^\rho(0)\,.
\label{eq_rel_loc_l}
\end{equation}

Therefore, we need to propose a more general scenario than the one considered above. 

\subsection{Relative locality from geometry: metric in phase space for two particles}
\label{sec:line_element}
Instead of regarding a line element for each particle separately, we can consider a line element for the whole phase space of both particles at the same time. While this could seem strange, note that the Lorentz transformation of the right particle in $\kappa$-Poincaré kinematics depends on the left momentum~\cite{Majid1994}, implying that one must regard the Lorentz transformations of the two-particle system as a transformation in the whole phase space
\begin{equation}
 {\cal J}^{\alpha\beta}\,=\,y^\mu  {\cal J}^{\alpha\beta}_\mu(p)+z^\mu   {\cal \tilde{J}}^{\alpha\beta}_\mu(p,q)\,.
\label{eq:transformation_ps}
\end{equation}

Then, we propose a line element in phase space of the form (for flat spacetime)
\begin{equation}
\mathcal{G}_2\,=\, G_{\mu\nu}(P) dX^A dX^B+G^{AB}(P) d P_A d P_B\,,
\label{eq:line_element_ps2}
\end{equation}
where $ G_{AB}(P) $ is an 8-dimensional metric
\begin{equation}
G_{AB}(P)\,=\,
\begin{pmatrix}
g^{LL}_{\mu\nu}(p,q) & g^{LR}_{\mu\nu}(p,q) \\
g^{RL}_{\mu\nu}(p,q) & g^{RR}_{\mu\nu}(p,q) 
\end{pmatrix}\,,
\label{eq:8-metric}
\end{equation}  
$X^A=(y^\mu,z^\mu)$, $P_A=(p_\mu,q_\mu)$, and $A$, $B$ run from $0$ to $7$. Explicitly, this line element can be written as
\begin{equation}
\begin{split}
\mathcal{G}_2\,=\,&g^{LL}_{\mu\nu}(p,q)dy^\mu dy^\nu +2 g^{LR}_{\mu\nu}(p,q)dy^\mu dz^\nu +g^{RR}_{\mu\nu}(p,q)dz^\mu dz^\nu+\\
&g_{LL}^{\mu\nu}(p,q)dp_\mu dp_\nu +2 g_{LR}^{\mu\nu}(p,q)dp_\mu dq_\nu +g_{RR}^{\mu\nu}(p,q)dq_\mu dq_\nu\,.
\end{split}
\end{equation}

In order to be a symmetric metric, their components satisfy 
\begin{equation}
g^{LL}_{\mu\nu}(p,q)\,=\,g^{LL}_{\nu\mu}(p,q)\,,\qquad g^{LR}_{\mu\nu}(p,q)\,=\,g^{RL}_{\nu\mu}(p,q)\,,\qquad g^{RR}_{\mu\nu}(p,q)\,=\,g^{RR}_{\nu\mu}(p,q)\,.
\end{equation}

Since we are considering only two particles in the initial state, and this is a classical model, we have also two particles in the final state, with phase-space coordinates $\left(u,k\right)$ and $\left(w,l\right)$. We want that the relative locality conditions~\eqref{eq_rel_loc_r},\eqref{eq_rel_loc_l} are satisfied. Then, we assume that the phase-space line element~\eqref{eq:line_element_ps2} is the same before and after the interaction
	\begin{equation}
\mathcal{G}_2\,=\, G_{AB}(P) dX^A dX^B+G^{\mu\nu}(P) d P_A d P_B\,=\, G_{AB}(K) dV^{A} dV^{B}+G^{AB}(K) d K_A d K_B \,,
\label{eq:line_element_rl}
\end{equation}
where $V^A=(u^\mu,w^\mu)$ and  $K_A=(k_\mu,l_\mu)$. In particular, we can define an intermediate stated as we did in Eq.~\eqref{eq:line_vertex}
	\begin{equation}
\begin{split}
 &G_{AB}(P) dX^A dX^B+G^{AB}(P) d P_A d P_B\,=\,2 g_{\mu\nu}\left(p\oplus q\right) d\xi^\mu d\xi^\nu+2 g^{\mu\nu}\left(p\oplus q\right)  d \left(p\oplus q\right)_\mu d\left(p\oplus q\right)_\nu \,=\,\\
&2g_{\mu\nu}\left(k\oplus l\right) d\xi^\mu d\xi^\nu+2g^{\mu\nu}\left(k\oplus l\right) d \left(k\oplus l\right)_\mu d\left(k\oplus l\right) _\nu\,=\, G_{AB}(K) dV^{A} dV^{B}+G^{AB}(K) d K_A d K_B\,.
\end{split}
\label{eq:line_element_vertex2}
\end{equation}
The factor 2 appears since we are asking for the two particles to have the same vertex of the interaction. Otherwise, we do not obtain the SR limit for which interactions are local. 

 This problem can be simplified if we use an 8-dimensional tetrad to depict the metric~\eqref{eq:8-metric}
\begin{equation}\Phi^A_B(p,q)\,=\,
\begin{pmatrix}
\varphi^{(L)\alpha}_{(L)\mu}(p,q) & \varphi^{(L)\alpha}_{(R)\mu} (p,q) \\
\varphi^{(R)\alpha}_{(L)\mu}(p,q) & \varphi^{(R)\alpha}_{(R)\mu}  (p,q)
\end{pmatrix}\,,
\label{eq:8-tetrad}
\end{equation}
such that 
	\begin{equation}
 G_{AB}(P) \,=\, \Phi^C_A(p,q)
\eta_{CD} \Phi^D_B(p,q)  \,,
\end{equation}
where 
	\begin{equation}
\eta_{CD} \,=\, \begin{pmatrix}
\eta_{\alpha\beta}&0\\
0& \eta_{\alpha\beta}
\end{pmatrix}  \,.
\end{equation}

It is easy to obtain the relationship between this tetrad and the components of the metric~\eqref{eq:8-metric}
\begin{equation}
\begin{split}
g^{LL}_{\mu\nu}(p,q)\,&=\,\varphi^{(L)\alpha}_{(L)\mu}(p,q)\eta_{\alpha\beta}\varphi^{(L)\beta}_{(L)\nu}(p,q)+\varphi^{(R)\alpha}_{(L)\mu}(p,q)\eta_{\alpha\beta}\varphi^{(R)\beta}_{(L)\nu}(p,q)\,,\\
g^{LR}_{\mu\nu}(p,q)\,&=\,g^{RL}_{\nu\mu}(p,q)\,=\,\varphi^{(L)\alpha}_{(L)\mu}(p,q)\eta_{\alpha\beta}\varphi^{(L)\beta}_{(R)\nu}(p,q)+\varphi^{(R)\alpha}_{(L)\mu}(p,q)\eta_{\alpha\beta}\varphi^{(R)\beta}_{(R)\nu}(p,q)\,,\\
g^{RR}_{\mu\nu}(p,q)\,&=\,\varphi^{(L)\alpha}_{(R)\mu}(p,q)\eta_{\alpha\beta}\varphi^{(L)\beta}_{(R)\nu}(p,q)+\varphi^{(R)\alpha}_{(R)\mu}(p,q)\eta_{\alpha\beta}\varphi^{(R)\beta}_{(R)\nu}(p,q)\,.
\end{split}
\label{eq:metric_tetrad}
\end{equation}

Therefore, the momentum part of line element part of Eq.~\eqref{eq:line_element_vertex2} will be satisfied if 
\begin{equation}
\begin{split}
\varphi^{\alpha}_{\mu}\left(p\oplus q\right)\,&=\,\frac{\partial \left(p\oplus q\right)_\mu}{\partial p_\nu}\varphi^{(L)\alpha}_{(L)\nu}(p,q) +\frac{\partial \left(p\oplus q\right)_\mu}{\partial q_\nu} \varphi^{(L)\alpha}_{(R)\nu} (p,q) \,,\\
\varphi^{\alpha}_{\mu}\left(p\oplus q\right)\,&=\,\frac{\partial \left(p\oplus q\right)_\mu}{\partial p_\nu}\varphi^{(R)\alpha}_{(L)\nu}(p,q) +\frac{\partial \left(p\oplus q\right)_\mu}{\partial q_\nu} \varphi^{(R)\alpha}_{(R)\nu} (p,q) \,.
\end{split}
\label{eq:varphi_L}
\end{equation}

Since we want to recover the metric for a single particle when there is only one momentum, we impose 
\begin{equation}
\varphi^{(L)\alpha}_{(L)\nu}(p,0)\,=\,\varphi^{\alpha}_{\mu}\left(p\right)\,,\qquad   \varphi^{(L)\alpha}_{(R)\nu} (0,q)\,=\, \varphi^{(R)\alpha}_{(L)\nu}(p,0)\,=\,0\,,\qquad   \varphi^{(R)\alpha}_{(R)\nu} (0,q)\,=\,\varphi^{\alpha}_{\mu}\left(q\right) \,,
\label{eq:varphi_L2}
\end{equation}
being $\varphi^{\alpha}_{\mu}\left(p\right)$ the (inverse of the) tetrad in momentum space 
\begin{equation}
 g_{\mu\nu}(k)\,=\,\varphi^{\alpha}_{\mu}(k)\eta_{\alpha\beta}\varphi^{\beta}_{\nu}(k)\,.
\end{equation}

Then, the desired relative locality condition~\eqref{eq_rel_loc_l} for the left particle can be obtained from the space-time part of the line element of Eq.~\eqref{eq:line_element_vertex2}. In terms of tetrads, one can obtain from Eq.~\eqref{eq:line_element_vertex2} 
\begin{equation}
\varphi^{\alpha}_{\nu}\left(p\oplus q\right)\,=\,\frac{\partial y^\mu }{\partial \xi^\nu}\varphi^{(L)\alpha}_{(L)\mu}(p,q) +\frac{\partial z^\mu }{\partial \xi^\nu} \varphi^{(L)\alpha}_{(R)\mu} (p,q)\,=\,\frac{\partial y^\mu }{\partial \xi^\nu}\varphi^{(R)\alpha}_{(L)\mu}(p,q) +\frac{\partial z^\mu }{\partial \xi^\nu} \varphi^{(R)\alpha}_{(R)\mu} (p,q) \,.
\label{eq:tetrad_tetrads}
\end{equation}
Therefore, due to the conditions~\eqref{eq:varphi_L} one can see that the previous equation is satisfied if
\begin{equation}
\frac{\partial y^\mu }{\partial \xi^\nu}\,=\,\frac{\partial \left(p\oplus q\right)_\nu}{\partial p_\mu}\,,\qquad\frac{\partial z^\mu }{\partial \xi^\nu}\,=\,\frac{\partial \left(p\oplus q\right)_\nu}{\partial q_\mu} \,,
\label{eq:rl_flat_coord}
\end{equation}
which is consistent with Eqs.~\eqref{eq_rel_loc_r},\eqref{eq_rel_loc_l}.

\section{Definition of spacetime from geometry}
\label{sec:st}
We have seen how to introduce a metric in the phase space of two particles in order to recover the relative locality principle. In this section we will see that this geometrical construction of such metric leads us to a new definition of spacetime.

\subsection{Features of noncommutative spacetime} 
We can now define from the space-time part of the line element~\eqref{eq:line_element_ps_flat} some new space-time coordinates as a function of the (inverse of the) momentum tetrad
\begin{equation}
\tilde{x}^\alpha\,=\,x^\mu \varphi^\alpha_\mu (k)\,. 
\label{eq:tilde_definition}
\end{equation}

We can consider now the propagation of a free massless particle in these noncommutative coordinates
\begin{equation}
ds^2\,=\, dx^\mu g_{\mu\nu}(k) dx^\nu \,=\,  dx^\mu \varphi_\mu^\alpha(k) \eta_{\alpha \beta}\varphi_\nu^\beta(k) dx^\nu\,=\,0 \,.
\end{equation}
Then, since  $\dot{k}=0$ along the trajectory, we have
\begin{equation}
 d\tilde{x}^\alpha \eta_{\alpha \beta} d\tilde{x}^\beta\,=\,0 \,.
\end{equation}

This means that in these coordinates there is an absence of a momentum dependence on times of flight for massless particles. This fact was previously pointed out in~\cite{Relancio:2020mpa}~\footnote{See also~\cite{Carmona:2018xwm,Carmona:2019oph,Carmona:2021gbg} for a different perspective of the same result.}. In that paper it is shown that, since ``physical'' distance (the one defined in terms of the noncommutative coordinates) traveled by a massless particle depends on its own momentum, there is a cancellation of effects, making that there is not an energy dependent time of arrival for photons. 

Now we can wonder what kind of noncommutativity can arise from this definition of spacetime. In particular, a very interesting example is the one in which they close the $\kappa$-Minkowski algebra~\cite{Carmona:2019fwf}
\begin{equation}
\left\{ \tilde{x}^\mu , \tilde{x}^\nu \right\}\,=\, \frac{1}{\Lambda}\left(n^\mu \tilde{x}^\nu- n^\nu \tilde{x}^\mu\right)\,,
\label{eq:tetrad_algebra}
\end{equation}
where $n^\mu$ is a fixed temporal vector $(1,0,0,0)$.

Note that this definition of spacetime differs from the one obtained in Hopf algebras, which in this geometrical setting would correspond to the generators of translations in momentum space (generators of the composition law)~\footnote{See~\cite{Carmona:2021gbg} for a more complete discussion. }.
\subsection{Noncommutative spacetime and locality of interactions} 
One can define a generalized spacetime, depending on all the phase-space variables of the two-particle system, in such a way that the space-time part of line element~\eqref{eq:line_element_ps2} can be rewritten as  
\begin{equation}
ds^2_2\,=\, d\tilde{y}^\alpha \eta_{\alpha\beta} d\tilde{y}^\beta+d\tilde{z}^\alpha \eta_{\alpha\beta} d\tilde{z}^\beta \,,
\label{eq:line_element_ps2_tildes}
\end{equation}
being
\begin{equation}
\tilde{y}^\alpha\,=\, y^\mu \varphi^{(L)\alpha}_{(L)\mu}(p,q) +z^\mu \varphi^{(L)\alpha}_{(R)\mu} (p,q)\,,\qquad \tilde{z}^\alpha\,=\, y^\mu \varphi^{(R)\alpha}_{(L)\mu}(p,q) +z^\mu \varphi^{(R)\alpha}_{(R)\mu} (p,q) \,,
\label{eq:st_tilde}
\end{equation}
and where we have used that momenta are constant since we are considering that the metric does not depend on the space-time coordinates. This defines some noncommutative coordinates in which interactions are local: from what we saw in the previous section, it is easy to check that 
\begin{equation}
\tilde{y}^\alpha(0)\,=\, \tilde{z}^\alpha(0)\,=\, \tilde{\xi}^\alpha(0)\,=\, \xi^\mu(0)\varphi^{\alpha}_{\mu}\left(p\oplus q\right) \,.
\end{equation}

This a more general case of the one considered in~\cite{Carmona:2019vsh}, in which the noncommutative space-time coordinates in which interactions are local were defined as   
\begin{equation}
\tilde{y}^\alpha\,=\, y^\mu \varphi^{\alpha}_{\mu}(p) +z^\mu \varphi^{(L)\alpha}_{(R)\mu} (q)\,,\qquad \tilde{z}^\alpha\,=\, y^\mu \varphi^{(R)\alpha}_{(L)\mu}(p) +z^\mu \varphi^{\alpha}_{\mu} (q) \,.
\label{eq:st_tilde_simple}
\end{equation}

In~\cite{Carmona:2019vsh} it was also pointed out that there is not an unequivocal way to define these $\varphi$'s functions given a relativistic deformed kinematics, even in the restricted case it was considered. Here, starting from a completely general set up, there are even more possible choices. In the next  subsection, we will give a way to select these functions in order to eliminate this ambiguity.

\subsection{New geometrical constraints to spacetime}
As commented in~\cite{Carmona:2019vsh,Carmona:2021gbg}, there are different ways to implement locality of interactions given a kinematics even in the one-particle system, despite imposing that they form a $\kappa$-Minkowski algebra. However, when considering this geometrical approach it is natural that, given a deformed relativistic kinematics, the $\varphi$ of one particle must be the (inverse of the) tetrad corresponding to the metric which has the Lorentz transformations of the one-particle system as the Lorentz isometries (as it was noted in~\cite{Carmona:2021gbg}).  

Moreover, we can follow this approach and select the $\varphi$'s functions of the two-particle system in such a way that the Lorentz transformations of the two-particle system are the isometries of the 8-dimensional metric. As we will see, this is a really strong condition that determines completely the non-commutative spacetime of a two-particle system. 
	
\section{Application to \texorpdfstring{$\kappa$}{k}-Poincaré kinematics}
\label{sec:kappa}
In this section, we apply our method for constructing a metric for a two-particle system to the kinematics of $\kappa$-Poincaré, and we see how this procedure can be generalized to three particles, being this easily generalizable to any number of particles. 
\subsection{Simple basis of  \texorpdfstring{$\kappa$}{k}-Poincaré}
\label{sec:simple_basis}
In~\cite{Carmona:2019vsh} a really simple basis of $\kappa$-Poincaré kinematics was found. The composition law reads
\be
\left(p\oplus q\right)_\mu \,=\, p_{\mu} + \left(1 - p_0/\Lambda\right) \,q_{\mu}\,.
\label{eq:dcl1}
\ee 
The Lorentz transformations for the left particle are
\begin{align}
  & {\cal J}^{ij}_0(k) \,=\, 0\,, \quad\quad\quad {\cal J}^{ij}_k(k) \,=\, \delta^j_k \, k_i - \delta^i_k \, k_j\,, \nonumber \\
  & {\cal J}^{0j}_0(k) \,=\, - k_j (1- k_0/\Lambda)\,, \quad\quad\quad {\cal J}^{0j}_k \,=\, \delta^j_k \left(-k_0 + \left(k_0^2-\vec{k}^2\right)/2\Lambda\right) + k_j k_k/\Lambda \,,
\label{LT1}
\end{align}  
while for the right one
\begin{align}
 {\cal \tilde{J}}_{0}^{\,0i}(p, q)\,=&\left(1- p_0/\Lambda\right) {\cal J}_0^{0i}(q)\,, \nonumber \\
 {\cal \tilde{J}}_{j}^{\,0i}(p, q)\,=&\left(1- p_0/\Lambda\right) {\cal J}_j^{0i}(q)- \,\left( \delta^i_j \vec{p}\cdot\vec{q}-p_j q_i \right)/\Lambda\,, \nonumber\\
 {\cal \tilde{J}}_{0}^{\,ij}(p, q)\,=&\, {\cal J}^{ij}_0(q)\,,  \quad\quad\quad
 {\cal \tilde{J}}_{k}^{\,ij}(p, q)\,=\, {\cal J}^{ij}_k(q)\,.
\label{calJ(2)}
\end{align}

  In order to have a relativity principle, the total momenta of the system for two observers must be related by a Lorentz transformation, and then the next equation holds
\begin{equation}
\left(p\oplus q\right)^\prime_\mu\,=\,\left(p^\prime\oplus \tilde{q}\right)_\mu\,,
\label{eq:rel_princ_comp}
\end{equation}
where
\be
p^\prime_\mu\,=\,p_\mu+\epsilon_{\alpha\beta}{\cal J}^{\alpha\beta}_\mu (p)\,,\qquad \tilde{q}_\mu\,=\,q_\mu+\epsilon_{\alpha\beta} {\cal \tilde{J}}^{\alpha\beta}_\mu(p,q)  \,.
\label{eq:tilde}
\ee
From Eq.~\eqref{eq:rel_princ_comp}, given the Lorentz transformations of Eq.~\eqref{eq:tilde}, one can check that the following expression holds~\cite{Carmona:2019fwf} 
\be
{\cal J}^{\alpha\beta}_\mu(p\oplus q) \,=\,\frac{\partial(p\oplus q)_\mu}{\partial p_\nu} {\cal J}^{\alpha\beta}_\nu(p) + \frac{\partial(p\oplus q)_\mu}{\partial q_\nu} {\cal \tilde{J}}^{\alpha\beta}_\nu(p,q)\,,
\label{eq_composition_Lorentz2}
\ee
having then the relativity principle present in a relativistic deformed kinematics.

The momentum metric with these isometries can be defined by a tetrad which satisfies Eq.~\eqref{eq:tetrad_algebra}
\be
\varphi^\alpha_\mu(k)\,=\,\delta^\alpha_\mu \left(1-k_0/\Lambda\right)\,.
\ee

One can obtain the Casimir as the squared distance of the metric constructed with the previous tetrad 
\be
C(k)\,=\,\,-\Lambda^2 \arcsec^2\left(\frac{2 \left(k_0-\Lambda\right)\Lambda}{-k_0^2+\vec{k}^2+2 \left(k_0-\Lambda\right)\Lambda}\right)\, .
\ee

As it was shown in~\cite{Carmona:2019vsh}, this basis can be obtained from the well-known  bicrossproduct basis~\cite{Majid1994} considering the change of momentum basis $k_\mu \to \hat{k}_\mu$ 
\be
k_i \,=\, \hat{k}_i\,, \quad\quad\quad (1 - k_0/\Lambda) \,=\, e^{- \hat{k}_0/\Lambda}\,,
\ee
being the hatted variables the one corresponding to the bicrossproduct basis. This particular basis will be used in the following due to its simplicity both on the kinematics and the corresponding metric. 

Due to the associativity of the composition law, there is a simple way to define the kinematics for more than two particles. In particular, for a system of three particles (which will be useful in the following) the generalization of the composition law~\eqref{eq:dcl1} is given by 
\be
\left(k\oplus\left(p\oplus q\right)\right)_\mu \,=\, k_{\mu} + \left(1 - k_0/\Lambda\right) \,\left(p\oplus q\right)_{\mu}\,,
\label{eq:composition_3}
\ee 
and the Lorentz transformations are
\be
k^\prime_\mu\,=\,k_\mu+\epsilon_{\alpha\beta}{\cal J}^{\alpha\beta}_\mu (k)\,,\qquad \tilde{p}_\mu\,=\,p_\mu+\epsilon_{\alpha\beta}{\cal \tilde{J}}^{\alpha\beta}_\mu (k,p)\,,\qquad \tilde{q}_\mu\,=\,q_\mu+\epsilon_{\alpha\beta}{\cal \tilde{J}}^{\alpha\beta}_\mu (k\oplus p,q)\,.
\label{eq:LT_3}
\ee 
It is easy to see that these transformations satisfy the generalization of Eq.~\eqref{eq_composition_Lorentz2} for three particles 
\be
{\cal J}^{\alpha\beta}_\mu\left(k\oplus\left(p\oplus q\right)\right) \,=\,\frac{\partial\left(k\oplus\left(p\oplus q\right)\right)_\mu}{\partial k_\nu} {\cal J}^{\alpha\beta}_\nu(k) +\frac{\partial\left(k\oplus\left(p\oplus q\right)\right)_\mu}{\partial p_\nu} {\cal \tilde{J}}^{\alpha\beta}_\nu(k,p) + \frac{\partial \left(k\oplus\left(p\oplus q\right)\right)_\mu}{\partial q_\nu} {\cal \tilde{J}}^{\alpha\beta}_\nu(k\oplus p,q)\,.
\label{eq_composition_Lorentz3}
\ee
 This can be generalized for any number of particles in an easy way. 

\subsection{Metric for the two-particle system}

As it was shown in~\cite{Carmona:2019fwf}, when considering a tetrad satisfying~\eqref{eq:tetrad_algebra} and the associative composition law of $\kappa$-Poincaré, the following relation between tetrad and composition law is satisfied 
\begin{equation}
\varphi^\mu_\nu(p \oplus q) \,=\,  \frac{\partial (p \oplus q)_\nu}{\partial q_\rho} \, \varphi_\rho^{\,\mu}(q)\,.
\label{eq:tetrad_composition}
\end{equation} 

Then, for the particular case of $\kappa$-Poincaré, the right particle satisfies the relative-locality condition~\eqref{eq_rel_loc_r}. This allows us to consider a simplified version of the 8-dimensional tetrad~\eqref{eq:8-tetrad}
\begin{equation}\Phi^\alpha_\mu(p,q)\,=\,
\begin{pmatrix}
\varphi^{(L)\alpha}_{(L)\mu}(p,q) & \varphi^{(L)\alpha}_{(R)\mu} (p,q) \\
0 & \varphi^{\alpha}_{\mu}  (q)
\end{pmatrix}\,,
\label{eq:8-tetrad2}
\end{equation}
having then that the second equation of~\eqref{eq:varphi_L} is automatically satisfied. 

Now we can impose that the Lorentz transformations~\eqref{eq:transformation_ps} must be an isometry of the metric 
\begin{equation}
 G_{AB}(P) dX^A dX^B+G^{AB}(P) d P_A d P_B\,=\, G_{AB}(P^\prime)dX^{\prime A} dX^{\prime B} +G^{AB}(P^\prime) d P^\prime_A d P^\prime_B\,,
\label{eq:line_element_LI}
\end{equation}
where $X^{\prime A}=(y^{\prime\mu},\tilde{z}^{\mu})$, being 
	\begin{equation}
y^{\prime\mu}\,=\,y^\mu+\epsilon_{\alpha\beta} \lbrace{y^\mu ,y^\rho  {\cal J}^{\alpha\beta}_\rho(p)+z^\rho   {\cal \tilde{J}}^{\alpha\beta}_\rho(p,q)\rbrace}\,,\qquad \tilde{z}^{\mu}\,=\,z^\mu+\epsilon_{\alpha\beta} \lbrace{z^\mu ,y^\rho  {\cal J}^{\alpha\beta}_\rho(p)+z^\rho   {\cal \tilde{J}}^{\alpha\beta}_\rho(p,q)\rbrace}\,,
\end{equation}
and  $P^\prime_A=(p^\prime_\mu,\tilde{q}_\mu)$, both defined in~\eqref{eq:tilde}.

Using this and the conditions~\eqref{eq:varphi_L}-\eqref{eq:varphi_L2}, we obtain the following expressions for the tetrad components of~\eqref{eq:8-tetrad2} when using the kinematics described at the beginning of this section
\begin{equation}
\begin{split}
\varphi^{(L)\alpha}_{(L)\mu}(p,q)\,&=\,\delta^\alpha_\mu \phi^L_1(p,q)+n^\alpha n_\mu\phi^L_2(p,q)+\frac{n^\alpha q_\mu}{\Lambda}\phi^L_3(p,q)+\frac{q^\alpha n_\mu}{\Lambda}\phi^L_4(p,q)+\frac{q^\alpha q_\mu}{\Lambda^2}\phi^L_5(p,q)\,,\\
\varphi^{(L)\alpha}_{(R)\mu}(p,q)\,&=\,\delta^\alpha_\mu \phi^R_1(p,q)+n^\alpha n_\mu\phi^R_2(p,q)+\frac{n^\alpha q_\mu}{\Lambda}\phi^R_3(p,q)+\frac{q^\alpha n_\mu}{\Lambda}\phi^R_4(p,q)+\frac{q^\alpha q_\mu}{\Lambda^2}\phi^R_5(p,q)\,,
\end{split}
\label{eq:tetrad_explicit}
\end{equation}
being
 \begin{equation}
\begin{split}
\phi^L_1(p,q)\,&=\,1-p_0/\Lambda\,,\quad \phi^L_2(p,q)\,=\,\frac{2\left(1-p_0/\Lambda\right)\left(q_0^2-\vec{q}^2\right)}{\vec{q}^2-\left(q_0-2\Lambda\right)^2}\,,\quad \phi^L_3(p,q)\,=\,-\frac{\phi^L_4(p,q)}{1-q_0/\Lambda}\,=\,-2\phi^L_5(p,q)\,=\,\frac{4\left(p_0-\Lambda\right)\Lambda}{\vec{q}^2-\left(q_0-2\Lambda\right)^2}\,,\\
\phi^R_1(q)\,&=\,-q_0/\Lambda\,,\quad \phi^R_2(q)\,=\,-2\phi^R_3(q)\,=\,-\frac{2\left(q_0^2-\vec{q}^2\right)}{\vec{q}^2-\left(q_0-2\Lambda\right)^2}\,,\quad \phi^R_4(q)\,=\,-2\phi^R_5(q)\,=\,\frac{4\left(q_0-\Lambda\right)\Lambda}{\vec{q}^2-\left(q_0-2\Lambda\right)^2}\,.
\end{split}
\end{equation}
Note that all the functions $ \phi^R$ depend only on the second momentum.

This completely determine the metric~\eqref{eq:8-metric} in the two-particle system
 \begin{equation}
\begin{split}
g^{LL}_{\mu\nu}(p)\,=\,&\left(1-p_0/\Lambda\right)^2\eta_{\mu\nu}\,,\quad g^{LR}_{\mu\nu}(p,q)\,=\,g^{LR}_{\nu\mu}(p,q)\,=\, 2 n_\mu  n_\nu  \frac{ (p_0-\Lambda ) (q_0^2-\vec{q}^2) (\Lambda -q_0)}{\Lambda ^2 (4 \Lambda  (\Lambda -q_0)+q_0^2-\vec{q}^2))}\\
&+\left(2q_\mu q_\nu-4n_\nu q_\mu \right)  \frac{ (p_0-\Lambda )(\Lambda -q_0)}{\Lambda ^2 (4 \Lambda  (\Lambda -q_0)+q_0^2-\vec{q}^2))}-n_\mu q_\nu   \frac{ (p_0-\Lambda )(4 q_0 \Lambda -3q_0^2-\vec{q}^2)}{\Lambda ^2 (4 \Lambda  (\Lambda -q_0)+q_0^2-\vec{q}^2))} +  \eta_{\mu\nu} \frac{q_0 \left(p_0-\Lambda\right)}{\Lambda^2}\,,\\
g^{RR}_{\mu\nu}(p,q)\,=\,& \left(1-\frac{2q_0}{\Lambda}-\frac{2q_0^2}{\Lambda^2}\right) \eta_{\mu\nu}+\left(n_\nu q_\mu+n_\mu q_\nu \right)\frac{2 \Lambda \left( q_0^2+\vec{q}^2\right)+q_0 \left(q_0^2-\vec{q}^2-4 \Lambda ^2\right)}{\Lambda ^2 (4 \Lambda  (\Lambda -q_0)+q_0^2-\vec{q}^2)}\\
&-4n_\mu  n_\nu \frac{ (q_0^2-\vec{q}^2)  (q_0-\Lambda )}{\Lambda  (4 \Lambda  (\Lambda -q_0)+q_0^2-\vec{q}^2)}+q_\mu q_\nu\frac{ \left(4 q_0 \Lambda -3q_0^2-\vec{q}^2\right)}{\Lambda ^2 (4 \Lambda  (\Lambda -q_0)+q_0^2-\vec{q}^2)}\,.
\end{split}
\end{equation}

Therefore using Eqs.~\eqref{eq:varphi_L} and~\eqref{eq:line_element_LI} one can define, without ambiguity, the spacetime of a two-particle system given that the noncommutativity for the one-particle system is $\kappa$-Minkowski. The crucial ingredient which eliminates this ambiguity in defining the spacetime of a two-particle system of~\cite{Carmona:2019vsh} is the strong condition~\eqref{eq:line_element_LI}, which imposes that the Lorentz transformations in the two-particle system are the isometries of the 8-dimensional metric.

\subsection{Metric for more than two particles}
For an interaction involving more than two particles in the initial state (and then also in the final one, since we are considering a classical model), the previous study can be generalized in a simple way when the composition law is associative. We will study the particular case of three particles the procedure,  being able to be generalized to any number of particles. 

Since we have to consider  the line element of three particles (with momenta $k,p,q$ and total momentum $(k\oplus p\oplus q)$), we will use a generalization of the tetrad~\eqref{eq:8-tetrad2}

\begin{equation}\Phi^\alpha_\mu(k,p,q)\,=\,
\begin{pmatrix}
\varphi^{(1)\alpha}_{(1)\mu}(k,p,q) & \varphi^{(1)\alpha}_{(2)\mu} (k,p,q) & \varphi^{(1)\alpha}_{(3)\mu} (k,p,q) \\
0 & \varphi^{(2)\alpha}_{(2)\mu} (p,q) & \varphi^{(2)\alpha}_{(3)\mu} (q) \\
0 &0 & \varphi^{(3)\alpha}_{(3)\mu} (q) 
\end{pmatrix}\,.
\label{eq:8-tetrad3}
\end{equation}
We take the tetrad of the two last particles $p,q$ to be independent of the momentum $k$ identifying 
\begin{equation}
\varphi^{(2)\alpha}_{(2)\mu} (p,q)\,=\,\varphi^{(L)\alpha}_{(L)\mu} (p,q)\,,\qquad\varphi^{(2)\alpha}_{(3)\mu} (q)\,=\,\varphi^{(L)\alpha}_{(R)\mu} (q)\,,\qquad\varphi^{(3)\alpha}_{(3)\mu} (q)\,=\,\varphi^{\alpha}_{\mu} (q)\,.
\end{equation}

Now Eq.~\eqref{eq:line_element_vertex2} is generalized to 
	\begin{equation}
G_{AB}(P) dX^A dX^B+G^{AB}(P) d P_A d P_B\,=\,3 g_{\mu\nu}\left(p\oplus q\right) d\xi^\mu d\xi^\nu+3 g^{\mu\nu}\left(p\oplus q\right)  d \left(p\oplus q\right)_\mu d\left(p\oplus q\right) ^\nu \,,
\label{eq:line_element_vertex3}
\end{equation}
where now the indexes $A$, $B$, run from $0$ to $11$ and $X^A$ and $P_A$ represent the phase-space coordinates of a system of three particles. The factor 3  appears since we are considering three particles involved in the interaction. This leads to the generalization of Eq.~\eqref{eq:varphi_L} for the tetrads for the first particle (with momentum $k$) 
\begin{equation}
\varphi^{\alpha}_{\mu}\left(k\oplus p\oplus q\right)\,=\,\frac{\partial \left(k\oplus p\oplus q\right)_\mu}{\partial k_\nu}\varphi^{(1)\alpha}_{(1)\nu}(k,p,q)+\frac{\partial \left(k\oplus p\oplus q\right)_\mu}{\partial p_\nu}\varphi^{(1)\alpha}_{(2)\nu}(k,p,q) +\frac{\partial \left(k\oplus p\oplus q\right)_\mu}{\partial q_\nu} \varphi^{(1)\alpha}_{(3)\nu} (k,p,q) \,.
\label{eq:varphi_1}
\end{equation}

Then, we can solve order by order the equations~\eqref{eq:varphi_1} and the generalization of~\eqref{eq:line_element_LI} for three particles for the kinematics described in Sec.~\ref{sec:simple_basis}  obtaining   (up to second order in the inverse of the high-energy scale) 
\begin{equation}
\begin{split}
\varphi^{(1)\alpha}_{(1)\mu}(k,p,q)\,=\,&\delta^\alpha_\mu \left(1-\frac{k_0}{\Lambda}\right)-\frac{p^\alpha p_\mu}{2\Lambda^2}+\frac{n^\alpha p_\mu}{\Lambda}\left(1-\frac{k_0+p_0+q_0}{\Lambda}\right) -\frac{p^\alpha n_\mu}{\Lambda}  \left(1-\frac{k_0}{\Lambda}\right)-\frac{p^\alpha q_\mu}{\Lambda^2}+\\
& n^\alpha n_\mu \frac{\left(\vec{p}+\vec{q}\right)^2-\left(p_0+q_0\right)^2}{2\Lambda^2}-\frac{p^\alpha q_\mu}{\Lambda} + \frac{n^\alpha q_\mu}{\Lambda}\left(1-\frac{k_0-q_0}{\Lambda}\right)- \frac{q^\alpha n_\mu}{\Lambda}\left(1-\frac{k_0+p_0}{\Lambda}\right)- \frac{q^\alpha q_\mu}{2\Lambda^2}\,,\\
\varphi^{(1)\alpha}_{(2)\mu}(p,q)\,=\,&-\delta^\alpha_\mu \left(\frac{p_0}{\Lambda}+\frac{p_0 q_0-\vec{p}\cdot \vec{q}}{\Lambda^2}\right)-\frac{p^\alpha p_\mu}{2\Lambda^2}+\frac{p^\alpha n_\mu}{\Lambda}+ n^\alpha n_\mu \frac{p_0^2+2 p_0q_0-\vec{p}^2-2\vec{p}\cdot \vec{q}}{2\Lambda^2}+\frac{p^\alpha q_\mu}{\Lambda} - \frac{n^\alpha q_\mu}{\Lambda}\frac{p_0}{\Lambda}\,,\\
\varphi^{(1)\alpha}_{(3)\mu}(p,q)\,=\,&\delta^\alpha_\mu \left(-\frac{q_0}{\Lambda}+\frac{p_0q_0-\vec{p}\cdot \vec{q}}{\Lambda^2}\right)+ n^\alpha n_\mu \frac{q_0^2-\vec{q}^2}{2\Lambda^2}-\frac{q^\alpha p_\mu}{\Lambda^2}+\frac{q^\alpha n_\mu}{\Lambda}-\frac{q^\alpha q_\mu}{2\Lambda^2}\,.
\end{split}
\label{eq:tetrad_3}
\end{equation}

We can see that there is not an easy way to extend the tetrad for the two-particle system to a generic multi-particle system. However, following this prescription, one can  generalize this construction for a system with any number of particles.

\section{Metric for a generic  relativistic deformed kinematics}
\label{sec:general}
In Sec.~\ref{sec:line_element} we proposed a systematic way to obtain the metric in phase space for a system of two particles. In this section, we are going to apply it to the most general kinematics at first order in the high-energy scale obtained in~\cite{Carmona:2012un}. 

The deformed dispersion relation compatible with rotational invariance as a function of the components of the momentum is parametrized by two adimensional coefficients $\alpha_1, \alpha_2$:
\begin{equation}
C(p)\,=\,p_0^2-\vec{p}^2+\frac{\alpha_1}{\Lambda}p_0^3+\frac{\alpha_2}{\Lambda}p_0\vec{p}^2=m^2\,,
\label{eq:DDR}
\end{equation}
while the deformed composition law is parametrized by five adimensional coefficients $\beta_1, \beta_2, \gamma_1, \gamma_2, \gamma_3$,
\begin{equation}
\left[p\oplus q\right]_0 \,=\, p_0 + q_0 + \frac{\beta_1}{\Lambda} \, p_0 q_0 + \frac{\beta_2}{\Lambda} \, \vec{p}\cdot\vec{q}\,, \,\,\,\,\,  \left[p \oplus q\right]_i \,=\, p_i + q_i + \frac{\gamma_1}{\Lambda} \, p_0 q_i + \frac{\gamma_2}{\Lambda} \, p_i q_0
+ \frac{\gamma_3}{\Lambda} \, \epsilon_{ijk} p_j q_k \,,
\label{eq:DCL}
\end{equation}
where $\epsilon_{ijk}$ is the Levi-Civita symbol. 

The most general form of the Lorentz transformations in the one-particle system is 
\begin{eqnarray}
\left[T(p)\right]_0 &\,=\,& p_0 + (\vec{p} \cdot \vec{\xi}) + \frac{\lambda_1}{\Lambda} \, p_0 (\vec{p} \cdot \vec{\xi})\,, \nonumber \\
\left[T(p)\right]_i &\,=\,& p_i + p_0 \xi_i + \frac{\lambda_2}{\Lambda} \, p_0^2 \xi_i + \frac{\lambda_3}{\Lambda} \, {\vec p}^{\,2} \xi_i + \frac{(\lambda_1 + 2\lambda_2 + 2\lambda_3)}{\Lambda} \, p_i ({\vec p} \cdot {\vec \xi}) \,,
\label{T-one}
\end{eqnarray}
where $\vec{\xi}$ is the vector parameter of the boost, and the $\lambda_i$ are dimensionless coefficients. 

The invariance of the dispersion relation under this transformation, $C(T(p))=C(p)$, requires the coefficients of the deformed dispersion relation to be a function of those of the boosts
\begin{equation}
\alpha_1 \,=-\,2(\lambda_1+\lambda_2+2\lambda_3)\,, \quad\quad \alpha_2\,=\,2(\lambda_1+2\lambda_2+3\lambda_3)\,.
\label{alphalambda}
\end{equation}

As we have mentioned previously, a modification in the transformations of the two-particle system is needed in order to have a relativity principle, making the deformed Lorentz transformations to depend on both momenta. Then, we are looking for a transformation such that $(p,q) \to (T^L_q(p),T^R_p(q))$, where
\begin{equation}
T^L_q(p) \,=\, T(p) + {\bar T}^L_q(p)\,, {\hskip 1cm} T^R_p(q) \,=\, T(q) + {\bar T}^R_p(q) \,.
\label{eq:boost2}
\end{equation}
When one considers the most general transformation in the two-particle system and imposes that they are Lorentz transformations and that they leave the deformed dispersion relation invariant, one finally finds:
\begin{eqnarray}
\left[{\bar T}^L_q(p)\right]_0 &\,=\,& \frac{\eta_1^L}{\Lambda} \, q_0 ({\vec p} \cdot {\vec \xi})  + \frac{\eta_2^L}{\Lambda} \, ({\vec p} \wedge {\vec q}) \cdot {\vec \xi}\,, \nonumber \\
\left[{\bar T}^L_q(p)\right]_i &\,=\,&  \frac{\eta_1^L}{\Lambda} \, p_0 q_0 \xi_i + \frac{\eta_2^L}{\Lambda}\left( \, q_0 \epsilon_{ijk} p_j \xi_k - p_0 \epsilon_{ijk} q_j \xi_k\right)+\frac{\eta_1^L}{\Lambda}\left(q_i ({\vec p} \cdot {\vec \xi})-({\vec p} \cdot {\vec q}) \xi_i \right) \,,
 \nonumber \\
\left[{\bar T}^R_p(q)\right]_0 &\,=\,& \frac{\eta_1^R}{\Lambda} \, p_0 ({\vec q} \cdot {\vec \xi}) + \frac{\eta_2^R}{\Lambda} \, ({\vec q} \wedge {\vec p}) \cdot {\vec \xi}\,, \nonumber \\
\left[{\bar T}^R_p(q)\right]_i &\,=\,& \frac{\eta_1^R}{\Lambda} \, q_0 p_0 \xi_i-\frac{\eta_2^R}{\Lambda} \left( p_0 \epsilon_{ijk} q_j \xi_k-q_0 \epsilon_{ijk} p_j \xi_k\right)+\frac{\eta_1^R}{\Lambda} \left(p_i ({\vec q} \cdot {\vec \xi}) - ({\vec q} \cdot {\vec p}) \xi_i\right) \,.\nonumber \\
\label{eq:gen2pboost}
\end{eqnarray}

Using Eq.~\eqref{eq_composition_Lorentz2} one finds the following relations between the coefficients of the composition law and those of the deformed Lorentz transformations 
\begin{alignat}{3}
\beta_1 &\,=\, 2 \,(\lambda_1 + \lambda_2 + 2\lambda_3)\,, \quad\quad & 
\beta_2 &\,=\, -2 \lambda_3 - \eta_1^L - \eta_1^R\,, \quad\quad & \label{betalambda}\\
\gamma_1 &\,=\, \lambda_1 + 2\lambda_2 + 2\lambda_3 - \eta_1^L\,, \quad\quad & 
\gamma_2 &\,=\, \lambda_1 + 2\lambda_2 + 2\lambda_3 - \eta_1^R\,, \quad\quad & \gamma_3 \,=\, \eta_2^L - \eta_2^R \,.
\label{gammalambda}
\end{alignat}

The most general tetrad for one particle making the corresponding metric invariant under the Lorentz transformation of one particle~\eqref{T-one} is 
\begin{equation}
\varphi^{\alpha}_{\mu}(k)\,=\,\delta^\alpha_\mu \left(1+\epsilon_1\frac{k_0}{\Lambda}\right)+\epsilon_2 \frac{n^\alpha k_\mu}{\Lambda}+\epsilon_3 \frac{k^\alpha n_\mu}{\Lambda}
 +\epsilon_4  k_0 \frac{n^\alpha n_\mu}{\Lambda}+\frac{\epsilon_5}{\Lambda} \epsilon^{\alpha\sigma\rho\nu}\eta_{\sigma\mu} k_\rho n_\nu\,,
\label{eq:tetrad_first_order_1}
\end{equation}
where
\begin{equation}
\epsilon_1\,=\,\lambda_1+2\lambda_2+2\lambda_3\,,\qquad \epsilon_3\,=\,-\epsilon_2+\lambda_1+2\lambda_2+4\lambda_3\,,\qquad \epsilon_4 \,=\,-2\lambda_2-2\lambda_3\,, 
\end{equation}
 and  $\epsilon_2$ and  $\epsilon_5$ are free parameters. 
 
We construct now the more general tetrad of two particles at first order in the high-energy scale 
\begin{equation}
\begin{split}
\varphi^{(M)\alpha}_{(N)\mu}(p,q)\,=\,&\delta^\alpha_\mu \left(\delta^{(M)}_{(N)}+\epsilon^{(M)}_{(N)1}\frac{p_0}{\Lambda}+\epsilon^{(M)}_{(N)2}\frac{q_0}{\Lambda}\right)+\epsilon^{(M)}_{(N)3}\frac{n^\alpha p_\mu}{\Lambda} +\epsilon^{(M)}_{(N)4}\frac{p^\alpha n_\mu}{\Lambda} +n^\alpha n_\mu \epsilon^{(M)}_{(N)5}\frac{p_0}{\Lambda}+\\
&\epsilon^{(M)}_{(N)6}\frac{n^\alpha q_\mu}{\Lambda}+\epsilon^{(M)}_{(N)7}\frac{q^\alpha n_\mu}{\Lambda}+n^\alpha n_\mu \epsilon^{(M)}_{(N)8}\frac{q_0}{\Lambda}+\epsilon^{(M)}_{(N)9}\epsilon^{\alpha\sigma\rho\nu}\eta_{\sigma\mu} p_\rho n_\nu+\epsilon^{(M)}_{(N)10}\epsilon^{\alpha\sigma\rho\nu}\eta_{\sigma\mu} q_\rho n_\nu\,,
\end{split}
\label{eq:tetrad_first_order_2}
\end{equation}
where $M,N$ can be $L,R$. 

Therefore, using the procedure explained in  Sec.~\ref{sec:line_element} one obtains the following expressions for the tetrads as a function of the coefficients of the Lorentz transformations of the two-particle system and the free coefficients of the tetrad of one particle 
\begin{equation}
\begin{split}
\epsilon^{(L)}_{(L)1}\,=\,&\epsilon_1\,,\quad \epsilon^{(L)}_{(L)2}\,=\,0\,,\quad \epsilon^{(L)}_{(L)3}\,=\,\epsilon_2\,,\quad \epsilon^{(L)}_{(L)4}\,=\,\epsilon_3\,,\quad \epsilon^{(L)}_{(L)5}\,=\,\epsilon_4\,,\\
\epsilon^{(L)}_{(L)6}\,=\,&\epsilon_2+\eta^L_1-\lambda_1-2\lambda_2-2\lambda_3\,,\quad \epsilon^{(L)}_{(L)7}\,=\,-\epsilon^{(L)}_{(L)6}\,,\quad \epsilon^{(L)}_{(L)8}\,=\,0\,,\quad \epsilon^{(L)}_{(L)9}\,=\,\epsilon_5\,,\quad \epsilon^{(L)}_{(L)10}\,=\,\epsilon_5+\eta^L_2\,,\\
\epsilon^{(L)}_{(R)1}\,=\,&0\,,\quad \epsilon^{(L)}_{(R)2}\,=\,\eta^R_1\,,\quad \epsilon^{(L)}_{(R)3}\,=\,0\,,\quad \epsilon^{(L)}_{(R)4}\,=\,0\,,\quad \epsilon^{(L)}_{(R)5}\,=\,0\,,\\
\epsilon^{(L)}_{(R)6}\,=\,&0\,,\quad \epsilon^{(L)}_{(R)7}\,=\,-\eta^R_1\,,\quad \epsilon^{(L)}_{(R)8}\,=\,0\,,\quad \epsilon^{(L)}_{(R)9}\,=\,\eta^R_2\,,\quad \epsilon^{(L)}_{(R)10}\,=\,0\,,\\
\epsilon^{(R)}_{(L)1}\,=\,&\eta^L_1\,,\quad \epsilon^{(R)}_{(L)2}\,=\,0\,,\quad \epsilon^{(R)}_{(L)3}\,=\,0\,,\quad \epsilon^{(R)}_{(L)4}\,=\,-\eta^L_1\,,\quad \epsilon^{(R)}_{(L)5}\,=\,0\,,\\
\epsilon^{(R)}_{(L)6}\,=\,&0\,,\quad \epsilon^{(R)}_{(L)7}\,=\,0\,,\quad \epsilon^{(R)}_{(L)8}\,=\,0\,,\quad \epsilon^{(R)}_{(L)9}\,=\,-\eta^L_2\,,\quad \epsilon^{(R)}_{(L)10}\,=\,0\,,\\
\epsilon^{(R)}_{(R)1}\,=\,&0\,,\quad \epsilon^{(R)}_{(R)2}\,=\,\epsilon_1\,,\quad \epsilon^{(R)}_{(R)3}\,=\,\epsilon_2+\eta^R_1-\lambda_1-2\lambda_2-2\lambda_3\,,\quad \epsilon^{(R)}_{(R)4}\,=\,- \epsilon^{(R)}_{(R)3}\,,\quad \epsilon^{(R)}_{(R)5}\,=\,0\,,\\
\epsilon^{(R)}_{(R)6}\,=\,&\epsilon_2\,,\quad \epsilon^{(R)}_{(R)7}\,=\,\epsilon_3\,,\quad \epsilon^{(R)}_{(R)8}\,=\,\epsilon_4\,,\quad \epsilon^{(R)}_{(R)9}\,=\,\epsilon_5\,,\quad \epsilon^{(R)}_{(R)10}\,=\,\epsilon_5+\eta^R_2\,.
\end{split}
\label{eq:coefficients_tetrad_first_order}
\end{equation}

It is important to note that with this construction we have two free parameters, $\epsilon_2$ and $\epsilon_5$. They can be fixed by imposing a particular algebra for the one-particle system. For example, the noncommutative coordinates defined in~\eqref{eq:tilde_definition} satisfy the $\kappa$-Minkowski algebra~\eqref{eq:tetrad_algebra} when $\epsilon_5=\eta^L_2=\eta^R_2=0$ and $\epsilon_2=1+\epsilon_1$.

Then, we see that the way in which we obtained the principle of relative locality can be used for any kinematics (in a generic way order by order in an expansion series on the high-energy scale): following the systematic way described in Sec.~\ref{sec:line_element} one can define a metric in phase space for a system of interacting particles with any kinematics. In particular, for the interesting cases of Snyder kinematics~\cite{Battisti:2010sr} and the so-called hybrid models~\cite{Meljanac:2009ej}.

Also, it is important to note that one cannot obtain the most general kinematics from the geometrical approach of~\cite{Carmona:2019fwf}. However, using this geometrical approach for a multi-particle system defining an eight dimensional momentum dependent metric, one is able to use any kind of composition law and Lorentz transformations satisfying the relativity principle.  This procedure allows us to define, for a given relativistic kinematics, the spacetime of a multi-particle system without ambiguity. Moreover, there is a crucial difference between both approaches. The construction of~\cite{Carmona:2019fwf} allows us to define in a simple way a family of relativistic kinematics (in which the generators of Lorentz transformations and translations close an algebra), with a deformed composition law, a deformed Casimir, and deformed Lorentz transformations in the two-particle system. The aim of the here presented approach is not to define a relativistic kinematics but, given any deformed kinematics (even the ones that are not obtained in~\cite{Carmona:2019fwf}), to construct the metric which implements the principle of relative locality, and then, the spacetime of a multi-particle system.

\section{Relative locality in curved spacetime}
\label{sec:rl_curved}

We can obtain the relative locality principle for a curved spacetime as we did in Sec.~\ref{sec:rl_flat} for the flat case. As showed in~\cite{Relancio:2020zok}, the isometries in momentum space when the metric depends also on the space-time coordinates are defined by a  modified composition ($\bar{\oplus}$) 
\be
(p \bar{\oplus} q)_\mu \,=\, e_\mu^\nu(\xi) (\bar{p} \oplus \bar{q})_\nu \,.
\label{eq:composition_cotangent}
\ee
where $p\rightarrow \bar{p}_\mu=\bar{e}_\mu^\nu(\xi) p_\nu$, $q\rightarrow \bar{q}_\mu=\bar{e}_\mu^\nu(\xi) q_\nu$, being  $\bar{e}^\mu_\lambda(x)$ is the inverse of the space-time tetrad $e^\lambda_\nu(x)$, satisfying $\bar{e}^\mu_\lambda(x)e^\lambda_\nu(x)=\delta^\mu_\nu$.  

Then, we apply this transformation on line element~\eqref{eq:line_element_ps}  
\begin{equation}
\mathcal{G}\,=\, g_{\mu\nu}(x,k) dx^\mu dx^\nu+g^{\mu\nu}(x,k) \delta k_\mu \delta k_\nu\,=\,g_{\mu\nu}(\xi ,\epsilon  \bar{\oplus}  k) d \xi ^\mu d\xi^\nu+g^{\mu\nu}(\xi ,\epsilon \bar{\oplus} k) \delta (\epsilon \bar{\oplus} k)_\mu \delta (\epsilon \bar{\oplus} k)_\nu\,,
\label{eq:line_element_ps_c_1} 
\end{equation} 
where $\delta k_\mu = dk_\mu-N_{\nu\mu}(x,k)dx^\nu$ and $\delta  (\epsilon \bar{\oplus} k)_\mu = d (\epsilon \bar{\oplus} k)_\mu-N_{\nu\mu}(\xi, (\epsilon\bar{\oplus} k))d\xi^\nu$.

The difficulty that arises here is that the composition law depends also on the space-time coordinates, making impossible to find a simple relation between the variables $(x,k)$ and $(\xi,(\epsilon \bar{\oplus} k))$ as we did in Sec.~\ref{sec:rl_flat}. The only way in which this relationship can be obtained is by considering a particular geometry, from which a differential equation involving the space-time coordinate will arise, leading to the analog version for a curved spacetime of~\eqref{eq:rl_flat_coord}. Also, one can realize that, since everything is defined through a metric in the cotangent bundle, all the previous results are invariant under space-time diffeomorphisms (see~\cite{Relancio:2020zok,Relancio:2020rys} for a discussion about diffeomorphisms in a cotangent bundle metric).   

It is important to note that, while the result in flat spacetime can be derived from the action~\eqref{S2}, the case for a curved spacetime has not a direct derivation. This is due to the fact that the composition law depends on the vertex of the interaction, which in the action is regarded as a Lagrange multiplier and not as a free parameter.

Notice also that this realization of relative locality in curved spacetime is completely different from the one obtained in~\cite{Cianfrani:2014fia}. In that paper it was considered an action and introduced some nonlocal variables (defined by the space-time tetrad). In this case, we are able to describe the relative locality principle in presence of a curvature on spacetime with the canonical variables, as it is done for the flat spacetime case. 

\section{Conclusions}
\label{sec:conclusions}
It is well known that a deformed relativistic kinematics can be obtained from a curved momentum space. This curved momentum space can be understood as a particular metric in the cotangent bundle geometry, leading to a momentum dependent space-time metric. 

Relative locality of interactions was understood from an action which involves the deformed composition law of momenta. Here, we proposed a novel way to obtain this principle from a geometrical point of view. Translations in momentum space depicted by a deformed composition law provokes modifications on the space-time coordinates when regarding the line-element in phase space. Then, since during interactions momenta change following this composition law, one can finally find the result of relative locality. This forces us to consider a metric for the phase space of two particles depending on all momenta involved in the interaction. 

From this metric, one can define some noncommutative coordinates in a multi-particle system recovering locality of interactions for all observers. While this construction was found to be ambiguous in other works, our geometrical perspective selects one particular implementation.

In this work we have studied how to construct a metric of two particles when considering a two-particle system for the $\kappa$-Poincaré kinematics in particular, showing how this work can be generalized for any relativistic kinematics. Moreover, we have shown that the procedure can be generalized for a system composed of more than two particles. This construction is not straightforward, so that the only way in which this can be done in general is in a series power expansion in the high-energy scale parametrizing the momentum dependence of the metric. 

We have also shown how to generalize the relative locality principle for a generic curved spacetime. This can be done thanks to our geometrical approach, since from an action there is not a simple generalization it in order to take into account a curvature on spacetime.  

We hope to study some phenomenological consequences of this geometrical implementation of relative locality  and go deeper in the notion of spacetime in future works.  

\section*{Acknowledgments}
The author would like to thank José Manuel Carmona, José Luis Cortés, and Stefano Liberati, for useful discussions. The author acknowledges support from the INFN Iniziativa Specifica GeoSymQFT. The author would also like to thank support from the COST Action CA18108. 

\end{document}